\documentstyle[12pt]{article}
\def\beq{\begin{equation}}
\def\ee{\end{equation}}
\def\bib#1{[{\ref{#1}}]}

\def\zbar{\overline{z}}

\begin{document}           

            \title{On classical string configurations}
 \author{B.G. Konopelchenko and G. Landolfi \\
{\em Dipartimento di Fisica, Universit\'{a} di Lecce, 
73100 Lecce, Italy}\\ and \\
{\em I.N.F.N., Sezione di Lecce, 73100 Lecce, Italy} } 

 \maketitle

      \begin{abstract}

 Equations which define classical configurations of strings in $R^3$ are 
 presented in a simple form. General properties as well as particular 
 classes of solutions of these equations are considered.
      \end{abstract}


Polyakov's theory of strings \bib{pol1}-\bib{pol2} has been discussed in a 
number of papers during the last ten years (see {\it e.g.} \bib{gross1}-
\bib{david1}). Various problems and features of this theory have been 
studied and analyzed. Stable classical configurations of strings is one of 
the important problems of the string theory. Their description, 
classification and analysis of their properties are essential for 
understanding classical, quasi-classical and other aspects of the string 
theory. Surprisingly, not many results concerning classical configurations 
of strings (world-sheets) 
have been obtained (see {\it e.g.} \bib{curt1}-\bib{david2}). The 
reason, obviously, lies in the complexity of the corresponding equations. So 
the description and analysis of the classical configurations of string 
is still of a great interest. 

In the present paper, we propose a new approach to the problem of classical 
configurations of strings in the three-dimensional Euclidean space. It is 
based on the generalized Weierstrass formulae which allows to construct any 
surface in $R^3$ starting with a system of two linear equations. Then we 
represent the Euler-Lagrange equation for the Nambu-Goto-Polyakov action in 
a simple form. Common solutions of this equation and above linear system 
provide surfaces in $R^3$ which describe classical configurations of 
strings (world-sheets). 
We discuss both general properties of the above system of 
equations as well as particular classes of solutions.  

Generalized Weierstrass formulae proposed in \bib{kon1} are the extension 
of well-known formulae of Weierstrass for minimal surfaces. One starts with 
the following system of two linear equations \bib{kon1}:
\beq
\left\{ 
\begin{array}{l}
\psi_{1 z}=p \psi_2 \\
\psi_{2 \zbar}=-p \psi_1
\end{array}
\right.
\label{eq:glp1}
\ee
where $\psi_1(z,\zbar), \psi_2(z,\zbar)$ are complex-valued functions, 
$p(z,\zbar)$ is a real-valued function and $z,\zbar$
are complex variables. Then one defines 
three real-valued functions $X_1, X_2, X_3$ as follows \bib{kon1}:
\begin{eqnarray} 
X_1 & + & i X_2  = 2i \int_{\Gamma} {\left( \overline{\psi}_1^2 dz'-
\overline{\psi}_2^2 d\zbar' \right)}  \nonumber \\
X_1 & - &  i X_2  = 2i \int_{\Gamma}{\left( \psi_2^2 dz'-
\psi_1^2 d\zbar' \right)}  \nonumber \\
X_3 &  = & - 2 \int_{\Gamma} { \left( \psi_2 \overline{\psi}_1 dz'
+\psi_1 \overline{\psi}_2 d\zbar' \right)} 
\label{eq:gweie}
\end{eqnarray}
where $\Gamma$ is an arbitrary path of integration in the complex plane
 and bar means 
complex conjugation. Due to (\ref{eq:glp1}), the integrals in
 (\ref{eq:gweie}) do not depend on the choice of $\Gamma$.
Finally, one treats $X_1, X_2, X_3$ as the coordinates of a surface in 
$R^3$ \bib{kon1}. The formulae (\ref{eq:gweie}) define the conformal 
immersion of a surface in $R^3$ since the corresponding first fundamental 
form is
\beq
d\ell^2=4 u^2  dz d\zbar
\label{eq:dl2}
\ee
where $u=|\psi_1|^2+|\psi_2|^2$. From (\ref{eq:dl2})
it is clear that $z,\zbar$ play the role of minimal lines 
($z=x+iy, \zbar=x-iy$). The Gaussian curvature $K$ and the mean 
curvature $H$ look like:
\begin{eqnarray}
K=-\frac{1}{u^2} \left( \log{u} \right)_{z\zbar} \;\; , 
\;\; H =\frac{p}{u}
\label{eq:KH}
\end{eqnarray}
The generalized Weierstrass formulae (\ref{eq:glp1})-(\ref{eq:gweie}) are the 
powerful tool for studying surfaces in $R^3$. In particular, any analytic 
surface in $R^3$ can be globally represented by the formulae 
(\ref{eq:glp1})-(\ref{eq:gweie}) 
\bib{tai1}. Note that another extension of 
the Weierstrass formulae for non-minimal surfaces has been proposed
 earlier 
in \bib{ken}. The equivalence of the generalized Weierstrass formulae 
(\ref{eq:glp1}) and those of \bib{ken} has been established in 
\bib{carr1}. However the form (\ref{eq:glp1}) is more convenient for 
many purposes.

The generalized Weierstrass formulae are also useful for an 
analysis of various quantities and problems of Polyakov string theory. 
Let us consider 
the standard Nambu-Goto-Polyakov action \bib{pol1}-\bib{pol2}
\beq
S=\mu_0 \int{d\Sigma} +\frac{1}{\alpha_0} \int{H^2 d\Sigma}
\label{eq:NGP}
\ee
where $d\Sigma$ is the area element. In terms of quantities 
$p, \psi_1, \psi_2$, from (\ref{eq:glp1})-(\ref{eq:gweie}) 
it has the following form \bib{carr1}:
\beq
S=4\mu_0 \int{\left( |\psi_1|^2+|\psi_2|^2 \right)^2 dx dy } +
\frac{4}{\alpha_0} \int{p^2 dx dy}
\ee
In the infrared region, where the second term in (\ref{eq:NGP}) dominates
\bib{pol2}, the action $S$ becomes quadratic in $p$ and the generating 
Polyakov's integral over all surfaces is reduced to a very simple one:
\beq
Z \sim \int{ \left[ d\sigma \right] e^{-\frac{4}{\alpha_0} \int{p^2 dx dy}} }
\label{eq:infrared}
\ee
The measure of integration $\left[ d\sigma \right]$ 
depends in general both on $p$ and 
$\psi_1,\psi_2$. We will discuss the functional (\ref{eq:infrared}) in 
a separate paper. Here, we will 
consider classical configurations of strings which are given 
by solutions of the Euler-Lagrange equation for the action (\ref{eq:NGP}). 

In generic coordinates the corresponding Euler-Lagrange equation has the 
form \bib{wil1}-\bib{ou1}:
\beq
\Delta H + 2H \left( H^2-K \right) -2\alpha_0\mu_0 H=0
\label{eq:eulag}
\ee
where $\Delta$ is the Laplace-Beltrami operator. If one chooses on a surface 
a conformal metric as in (\ref{eq:dl2}), then 
\beq
\Delta H = \frac{1}{u^2} H_{z\zbar}
\ee
and in terms of the variables $\varphi=H^{-1}$ and $p=u / \varphi$ equation 
(\ref{eq:eulag}) takes the form:
\beq
\varphi_{z\zbar} + \left[ 2p^2+ \left(\log{p^2}\right)_{z\zbar} \right] 
\varphi-2\alpha_0\mu_0 p^2 \varphi^3=0
\label{eq:eulagconf}
\ee
Note that the Euler-Lagrange equation (\ref{eq:eulag}) has the form  
(\ref{eq:eulagconf}) in the conformal metric independently of any use of 
the generalized Weierstrass formulae (\ref{eq:glp1})-(\ref{eq:gweie}). But if 
one starts with the Weierstrass formulae (\ref{eq:glp1})-(\ref{eq:gweie}), 
then the quantity $p$ in (\ref{eq:eulagconf}) is exactly the coefficient 
$p$ in equation (\ref{eq:glp1}) and $\varphi=\left( |\psi_1|^2 +
 |\psi_2|^2 \right)/p$. Thus equations (\ref{eq:glp1}), (\ref{eq:eulagconf}) 
form a system of equations which completely defines the 
classical configurations of strings in $R^3$ via the formulae 
(\ref{eq:gweie}). Any surface in $R^3$ is constructable by the formulae 
(\ref{eq:glp1})-(\ref{eq:gweie}). Those which obey in addition equation 
(\ref{eq:eulagconf}) represent classical configurations of strings. The 
system of 
equation (\ref{eq:glp1}), (\ref{eq:eulagconf}) is much simpler than those in 
terms of coordinates $X_{1},X_2,X_3$ derived and used in \bib{curt1}-
\bib{pis1}. Equation (\ref{eq:eulagconf}) itself (without the generalized 
Weierstrass formulae (\ref{eq:glp1})-(\ref{eq:gweie})) defines possible 
candidates to the classical configurations of strings. So it is of the great 
interest and importance itself. We will present below 
some classes of its solutions. 

We will consider here three particular cases: {\it 1)} constant mean 
curvature; {\it 2)} constant mean curvature density; {\it 3)} the case
$\mu_0=0$. The simplest solution of equation (\ref{eq:eulagconf}) 
corresponds to the constant mean curvature, {\it i.e.} to 
$\varphi=\varphi_0=$
constant. In this case equation (\ref{eq:eulagconf}) is reduced to the 
Liouville equation for $\theta=2\log{p}$:
\beq
\theta_{z\zbar}+\beta e^{\theta}=0
\label{eq:liou}
\ee
where $\beta=2\varphi_0-2\alpha_0\mu_0\varphi_0^3$. The general solution of 
equation (\ref{eq:liou}) is of the form (see {\it e.g.} \bib{dub1})
\beq
\exp{\theta}=
\frac{A_z \overline{A}_{\zbar}}{ \left( |A|^2+\beta/2 \right)^2}=p^2
\label{eq:solliou}
\ee
while for $\beta=0$ one has $p^2=A(z) \overline{A}(\zbar)$ 
where $A(z)$ is an arbitrary analytic function. 
So the formula (\ref{eq:solliou}) gives us two cylindrical 
surfaces ($\beta=0$) with $H_0=\pm \sqrt{\alpha_0 \mu_0}$. 
At $H>\sqrt{\alpha_0 \mu_0}$ and 
$-\sqrt{\alpha_0 \mu_0}<H<0$ surfaces are 
spherical ($\beta>0$ and, hence, $K>0$) while for 
$H<-\sqrt{\alpha_0 \mu_0}$ and $0<H<\sqrt{\alpha_0 \mu_0}$ one 
has pseudo-spherical surfaces ($\beta <0$ and $K<0$). 

Now we consider as second particular case $H\sqrt{g}=const.$, namely
$p=p_0=const.$ Equation 
(\ref{eq:eulagconf}) is now 
\beq
\varphi_{z\zbar}+2p_0^2 \varphi-2\alpha_0\mu_0 p_0^2 \varphi^3=0
\label{eq:eulag2}
\ee
In the simplest case $\mu_0=0$, we have the linear equation  
\beq
\varphi_{xx}+\varphi_{yy}+8p_0^2 \varphi=0
\label{eq:eulag3}
\ee
General solution and classes of particular solutions of this equation are 
easily available. For the square domain $0 \leq x,y \leq \pi$ and vanishing 
$\varphi$ at the boundary, one has the well-known family of solutions
\beq
\varphi_{nm}=\frac{1}{H_{nm}}=A_{nm} \sin{nx} \;  \sin{my}
\ee
where $n,m=1,2,\dots$, $A_{nm}$ are constants and $p_0$ takes discrete 
values $p_0^2=(n^2+m^2)/8$. So the corresponding closed surface has
the following first fundamental form, string action and total area
\begin{eqnarray}
&d\ell^2&=\frac{n^2+m^2}{2} 
\left( A_{nm} \sin{nx} \;  \sin{mx} \right)^2 (dx^2+dy^2) \\
&S_{nm}&=\frac{(n^2+m^2)}{2\alpha_0}
\label{eq:sbox} \\
&\Sigma_{nm}&=\frac{\pi^2(n^2+m^2)A_{nm}^2}{2} 
\end{eqnarray}
respectively. Equation (\ref{eq:eulag2}) is well-known in various fields 
of physics. It has the famous kink-antikink solution \bib{raja1}:
\beq
\varphi=\pm \left( \alpha_0 \mu_0 \right)^{-1/2}
\tanh{ \left[p_0 \left( e^{ia} z+ e^{-ia} \zbar +b \right) \right]}
\label{eq:kink}
\ee
where $a,b$ are arbitrary real constants. It must be noted that the mean
curvature $H=\varphi^{-1}$ is singular along the line 
$2x\cos{a}-2y\sin{a}+b=0$.
Let us consider now the case when $\varphi$ depends only on
$x=(z+\zbar)/2$. Equation (\ref{eq:eulag2}) is equivalent to 
\beq
\varphi_x^2+8p_0^2 \varphi^2-4\alpha_0 \mu_0 p_0^2 \varphi^4=const.
\ee
Solution of this equation are given by standard elliptic functions. 
Equation (\ref{eq:eulag2}) arises also in a more general case, namely 
when $\left( \log{p} \right)_{z\zbar}=0$. Since then $p^2=A(z) 
\overline{A}(\zbar)$ -where $A(z)$ is an arbitrary analytic function- 
equation (\ref{eq:eulagconf}) is reduced to (\ref{eq:eulag2}) after the 
change of variable $z \rightarrow A(z)$. 

Among the case when both $p \neq const$ and $\varphi \neq const$ we 
mention one with 
$p^2= 2 A_z \overline{A}_{\zbar} \left( A+\overline{A} \right)^{-2}$ 
where $A(z)$ is an arbitrary analytic function. Changing $z \rightarrow 
\xi=A(z)$, one gets from (\ref{eq:eulagconf}) the equation
\beq
(\xi +\overline{\xi})^2 \varphi_{\xi \overline{\xi}} +
6\varphi - 4 \alpha_0\mu_0 \varphi^3=0 
\ee
In the pure Polyakov case $\mu_0=0$ 
this equation becomes linear and general solution is of the form 
(see \bib{fors1}) 
$\varphi=\left( \partial_{\xi} - \frac{3}{\xi+\overline{\xi}}
 \right)^2 B(\xi) + c.c.$ 
where $B(\xi)$ is an arbitrary analytic function.

Among the different possible constraints on $p$
and $\varphi$ there is one of importance in string theory. It is the gauge 
of constant mean density $H \sqrt{g}=1$ in which the Virasoro symmetry is 
easily revealed (see {\it e.g.} \bib{gross1}). In our variables it is the 
constraint $p^2\varphi=1$. In this gauge one has 
$S=4 \int{ \left( \mu_0\varphi+ \alpha_0^{-1} \varphi^{-1}
\right) dx dy}$ 
and equation (\ref{eq:eulagconf}) looks like
\beq
\varphi_z \varphi_{\zbar} + 2 \varphi -2 \alpha_0 \mu_0 \varphi^3=0
\label{eq:HG1}
\ee
In the one-dimensional case (say $\varphi_y=0$), setting
$\varphi=\left( 2\alpha_0\mu_0 \right)^{-1} \wp$ this equation is reduced to 
\beq
\wp_x^2=4\wp^3-16\alpha_0\mu_0 \wp
\ee
that is the standard equation for the Weierstrass elliptic function with 
the invariants $g_2=16 \alpha_0 \mu_0$ and $g_3=0$.
Equation (\ref{eq:HG1}) is also solvable by the elliptic change of the 
dependent variable and then by the use of the method of characteristic. For
example, this procedure gives
\beq
\varphi=\wp \left( c+\sqrt{2\alpha_0\mu_0} \sqrt{(x-a)^2+(y-b)^2}
\right)
\ee
where $c,a$ and $b$ are constants and now 
$\wp$ is the Weierstrass elliptic 
function with the invariants $g_2=4/\alpha_0\mu_0$ and $g_3=0$.

Now we will consider our third particular case $\mu_0=0$, when equation 
(\ref{eq:eulagconf}) is reduced to the linear equation
\beq
\varphi_{z\zbar}+V(z,\zbar) \varphi=0
\label{eq:eulaglin}
\ee
where $V=2 \left[ \left( \log{p} \right)_{z\zbar} + p^2 \right]$.
For periodic functions $p$ we have $\int{H^2 d\Sigma}=4\int{H^2 dx 
dy}$. So the action (\ref{eq:NGP}) in this case is completely defined by 
the {\it potential} $V(z,\zbar)$. If $V=const$, 
we have an infinite family of solutions for equation 
(\ref{eq:eulaglin}). 
To define $p$ one has to solve the equation $\theta_{z\zbar}+
2 \exp{\theta}=V=const.$, where $\theta=p^2$. The existence of a 
wide class of periodic solutions of this 
equation was claimed in \bib{ves1}. Using those 
results and solution of equation (\ref{eq:eulag3})
 presented above, one can construct the 
variety of closed surfaces  with the same value of action
$S=2\pi^2 V/ \alpha_0$.

In the one-dimensional case $\varphi=\varphi(x)$ one can use the known 
results about the one-dimensional Schr\"{o}dinger equation 
$-\varphi_{xx}+Q(x)\varphi=E\varphi$. There are several solvable cases. 
One of them is provided by the potential $Q=\frac{2}{\sin{x}^2}$ 
(Sutherland model \bib{suth1}). Taking the periodic wave function
$\varphi=\frac{2\sqrt{2}}{\sin{x}}$ associated with the unphysical 
value of the energy $E=1$, one has $V=\frac{1}{4}-\frac{1}{2\sin{x}^2}$. 
Calculating $p^2$, one gets 
$p=\frac{\sin{x}}{2\sqrt{2} \left( \sqrt{2}-\sin{x} \right) }$. 
Thus $H=\frac{\sin{x}}{2\sqrt{2}}$ and $u=\left( \sqrt{2}
-\sin{x} \right)^{-2}$. The corresponding surface is nothing but the 
Clifford torus (see \bib{wil1}). Here we would like to emphasize its 
connection with the solvable (Sutherland) case of equation 
(\ref{eq:eulaglin}). Other solvable potentials in (\ref{eq:eulaglin}) 
may provide remarkable surfaces too. Nevertheless, 
it must be pointed out that the 
requirement of a physical interpretation is crucial and 
we must not forget that we are dealing with real surfaces. 
For instance, for $Q=-\frac{2}{\cosh{x}^2}$ (Bargmann potential) 
and $E=1$, one has a real mean curvature 
$H=\frac{\cosh{x}}{2\sqrt{2}}$ and pure imaginary 
$p=\frac{i\cosh{x}}{2\sqrt{2} \left( \sqrt{2}+\cosh{x}\right)}$. 
Consequently, the metric is negatively defined: 
$u^2=-\left( \sqrt{2}+\cosh{x}\right)^{-2}$.
Here, the reality of $p$ is the problem. 

A variety of exact solutions for the two-dimensional Schr\"{o}dinger 
equation and, hence, equation (\ref{eq:eulaglin}) is available too. 
They have been 
constructed by the inverse-spectral transform method (see {\it e.g.} the 
review \bib{kon2}) and are expressed in terms of the Prym 
$\theta$-function. 

Now we will discuss some examples of solutions of system of equations
(\ref{eq:glp1}), (\ref{eq:eulagconf}) and the corresponding surfaces 
given by formulae (\ref{eq:glp1})-(\ref{eq:gweie}).
The first example is given by 
\begin{eqnarray}
p=\frac{1}{4}, \;\; \psi_1=\psi_2=\frac{\sqrt{R}}{2} \exp{\frac{iy}{2}}
\end{eqnarray}
So the potential $V=1/8$ is constant and we get a cylinder of radius $R$:
\begin{eqnarray}
X_1=-R \sin{y}, \;\; X_2=-R \cos{y}, \;\; X_3=-R x
\end{eqnarray}
The second case is
\begin{eqnarray}
& p & =0 \nonumber \\
& \psi_1 & =\frac{1}{2} \exp{\left( \frac{iy-x}{2} \right)}, \;\;
\psi_2 = \frac{1}{2} \exp{\frac{iy+x}{2}}
\end{eqnarray}
and
\begin{eqnarray}
X_1=-\sin{y} \; \cosh{x},\;\; X_2=-\cos{y} \; \cosh{x}, \;\;X_3=-x
\end{eqnarray}
namely the catenoid.
The third case is
\begin{eqnarray}
& p & =\frac{1}{2\cosh{x}} \nonumber \\
& \psi_1 & =\frac{1}{2\cosh{x}} \exp{\left( \frac{iy+x}{2} \right)}, \;\; 
\psi_2 = \frac{1}{2\cosh{x}} \exp{\frac{iy-x}{2}}
\end{eqnarray}
and 
\begin{eqnarray}
X_1=-\frac{\sin{y}}{\cosh{x}},\;\; X_2=-\frac{\cos{y}}{\cosh{x}},\;
\;X_3=-\tanh{x}
\end{eqnarray}
that is the unit sphere $X_1^2+X_2^2+X_3^2=1$. Also in this case 
$V=0$.

Fourth example : the Clifford torus discussed above (which can be obtained 
by rotation of the circle of radius $1$ with the centre at $(\sqrt{2},0)$ 
corresponds to
\begin{eqnarray}
& p & =\frac{\sin{x}}{2 \sqrt{2} 
\left( \sqrt{2}-\sin{x} \right)} \nonumber \\
& \psi_1 & =\frac{\sqrt{\left( \sqrt{2} 
-\sin{x} -\cos{x} \right)}}{2 \left( \sqrt{2} 
- \sin{x} \right)} e^{\frac{iy}{2}}, \;\; 
\psi_2 =\frac{\sqrt{\left( \sqrt{2} - 
\sin{x} +\cos{x} \right)}}{2 \left( \sqrt{2} 
- \sin{x} \right)} e^{\frac{iy}{2}} \nonumber
\end{eqnarray}
and
\begin{eqnarray}
X_1=-\frac{\sin{y}}{\sqrt{2}-\sin{x}} ,\;\;
X_2=-\frac{\cos{y}}{\sqrt{2}-\sin{x}} ,\;\;
X_3=-\frac{\cos{x}}{\sqrt{2}-\sin{x}}
\end{eqnarray}
The potential is $V=\frac{1}{4}-\frac{1}{2\sin^2{x}}$. 
For the Clifford torus $\int{H^2 d\Sigma}=2\pi^2$. 
According to the conjecture of 
Wilmore \bib{wil1}, the Clifford torus provides 
the minimum for the functional 
$\int{H^2 d\Sigma}$ within a class of compact surfaces of genus $1$. 

And finally, the fifth example which is associated to the potential 
$$ V(x)= \frac{R^4-8R(R-\sin(f))}{8 [R-2\sin(f)]^2 } $$
Here $f_x(x)=R-\sin[f(x)]$, {\it i.e.}
$$f(x)=2\arctan{h(x)}$$
with $ h(x)=\frac{1}{R}+
\sqrt{1-\frac{1}{R^2}} \tan\left[ \sqrt{R^2-1} (x-C_1)/2 \right]
$
from which follows that
$$\sin[f(x)]=\frac{2h(x)}{1+h^2(x)} $$
The corresponding surface, defined by
\begin{eqnarray}
 p=\frac{R-2\sin[f(x)]}{4} ,\;\;
\psi_1=\sqrt{\frac{f-f_{x}}{4}} e^{\frac{iy}{2}} ,\;\;
\psi_2=\sqrt{\frac{f+f_{x}}{4}} e^{\frac{iy}{2}}
\end{eqnarray}
and, consequently, 
\begin{eqnarray}
X_1=-f(x) \cos{y} ,\;\; X_2=-f(x) \sin{y} ,\;\;
X_3 =-\cos{f(x)}
\end{eqnarray}
represents an extension of the Clifford torus \bib{tai2}.

Other analytic and numerical solution of equations 
(\ref{eq:glp1})-(\ref{eq:eulagconf}) and corresponding 
classical configurations of strings will be discussed elsewhere. 

 \hfill
 \hfill
 \hfill
 \hfill


   \begin{centerline} 
   {\bf REFERENCES}
   \end{centerline}

   \begin{enumerate}

  \item \label{pol1}
     A.M. Polyakov, Phys. Lett. {\bf B103}, 207, (1981).
  \item \label{pol2}
     A.M. Polyakov, Nucl. Phys. {\bf B268}, 406, (1986).
  \item \label{gross1}
     D.I. Gross, T. Piran and S. Weinberg Eds., {\it Two dimensional 
quantum gravity and random surfaces}, World Scientific, Singapore, (1992).
  \item \label{david1}
  F. David, P. Ginsparg and Y. Zinn-Justin Eds., {\it Fluctuating 
geometries in statistical mechanics and field theory}, Elsevier Science, 
Amsterdam, (1996).  
  \item \label{curt1}
  T.L. Curtright, G.I. Chandour, C.B. Thorn and C.K. Zachos, Phys. Rev. 
Letts, {\bf 57}, 799, (1986).
  \item \label{curt2}
  T.L. Curtright, G.I. Chandour and C.K. Zachos, Phys. Rev. {\bf D34}, 
3811, (1986).
  \item \label{bra1}
  E.Braaten, R.D. Pisarski and Sze-Man Tse, Phys. Rev. Letts. {\bf 58}, 93, 
(1987).
  \item \label{bra2}
  E. Braaten and C.K. Zachos, Phys. Rev {\bf D35}, 1512, (1987).
  \item \label{pis1}
  R. Pisarski, Phys. Rev. Letts. {\bf 58}, 1300, (1987).
  \item \label{ole1}
  P. Olesen and Sung-Kil Yang, Nucl. Phys. {\bf B283}, 73, (1987). 
  \item \label{david2}
  F. David and E. Guitter, Nucl. Phys. {\bf B295}, 332, (1988).
  \item \label{kon1}
   B.G.Konopelchenko, Stud. Appl. Math. {\bf 96}, 9, (1996).
  \item \label{tai1}
  I.A. Taimanov, preprint dg-ga/9511005.
  \item \label{ken}
  K. Kenmotsu, Math. Ann. {\bf 245}, 89, (1979).
  \item \label{carr1}
  R. Carroll and B.G. Konopelchenko, Int. J. Mod. Phys. {\bf A11}, 1183, 
 (1996).
  \item \label{wil1}
  T.I. Wilmore, {\it Riemannian Geometry}, Clatendon Press, Oxford (1993)
  \item \label{ou1}
  Ou-Yang Zhong-can and W. Helfrich, Phys. Rev. {\bf A39}, 5280, (1989).
  \item \label{dub1}
  B.A. Dubrovin, S.P. Novikov and A.T. Fomenko, {\it Modern Geometry}, 
Nauka Moscow (1979).
  \item \label{raja1}
  R. Rajaraman, {\it Solitons and Istantons}, North-Holland P.C., Amsterdam 
(1982).
  \item \label{fors1}
  A.R. Forsyth, {\it Theory of differential equations}, part IV, v VI, 
Dover I. (1906).
  \item \label{bat1}
  H. Bateman, {\it Higher transcendental functions}, v.3, Mc Graw-Hill B.C., 
New York (1955).
  \item \label{ves1}
  A.P. Veselov and S.P. Novikov, Usp. Mat. Nauk, {\bf 50}, n. 6,171 (1995) 
  \item \label{olsh1}
  M.A. Olshanetsky and A.M. Perelomov, Phys. Rep., {\bf 71}, 314, (1981).
  \item \label{suth1}
  B. Sutherland, Phys. Rev., {\bf A5}, 1372, (1972).
  \item \label{kon2}
  B.G. Konopelchenko, {\it Introduction to multidimensional integrable 
equations}, Plenum Press, New-York (1992).
  \item \label{tai2}
  I. A. Taimanov, {\it Surfaces of revolution in terms of solitons}, 
 dg-ga/9610012.

   \end{enumerate}

\end{document}